\documentstyle[prl,aps,preprint]{revtex}
\begin{document}
\draft
\title{GAUGE INVARIANT EFFECTIVE STRESS-ENERGY TENSORS FOR
GRAVITATIONAL WAVES}
\author{Paul R. Anderson}
\address{Department of Physics, Wake Forest University, P.O. Box
7507, Winston-Salem, NC  27109}

\maketitle
\begin{abstract}
It is shown that if a generalized definition of gauge invariance
is used, gauge invariant effective stress-energy tensors for
gravitational waves and other gravitational perturbations can be
defined in a much larger variety of circumstances than has
previously been possible.  In particular it is no longer
necessary to average the stress-energy tensor over a region of
spacetime which is larger in scale than the wavelengths of the
waves and it is no longer necessary to restrict attention to high
frequency gravitational waves.  
\end{abstract}

     It has long been known that gravitational waves can have an
effective stress-energy associated with them which alters the
background spacetime on which they propagate.  One important
example is the two body problem in general relativity where the
emission of gravitational radiation causes the orbit to decay. 
This effect has been observed for the binary pulsar\cite{pulsar}. 
A second example is the gravitational geon solution found by
Brill and Hartle\cite{BH}, BH.  The geon consists of high
frequency gravitational waves confined to a thin spherical shell
by the background geometry which they create.  Outside the shell
the geometry is the same as that outside of a static star.  

To quantify the backreaction effects of gravitational waves it is
necessary to define an effective 
stress-energy tensor for the waves.  This was first done by BH
and later by Isaacson\cite{I} for high frequency gravitational
waves in a vacuum.  Burnett\cite{B} also defined an effective
stress-energy tensor for high frequency waves both in a vacuum
and in spacetimes containing classical matter.  Efroimsky\cite{E}
extended Isaacson's definition to include lower frequency waves
and spacetimes containing classical matter.  In each case either
some sort of averaging procedure or a procedure that gives
similar results to an averaging procedure was used.

An important property of the wave equation and 
stress-energy tensor for gravitational waves is gauge invariance. 
The usual gauge transformations used for gravitational waves are
related to infinitesimal coordinate transformations and gauge
invariance implies invariance under these transformations. 
Isaacson showed that, in general, the wave equation for
gravitational waves is approximately gauge invariant only for
high frequency waves.  He also showed that for high frequency
waves the effective stress-energy tensor is gauge invariant to
leading order only if it is averaged over a region of spacetime
whose scale is large compared to the wavelengths of the waves. 
In other cases such as low frequency gravitational waves or the
gravitational geon where the averaging is over time, either the
wave equation, the stress-energy tensor, or both are not gauge
invariant.

In this paper it is shown that if a generalized gauge
transformation which is related to arbitrary coordinate
transformations is used, then it is possible to define gauge
invariant effective stress-energy tensors for gravitational waves
and other gravitational perturbations in virtually all situations
of interest.  This includes the cases of both low and high
frequency gravitational waves.  The stress-energy tensor can be
averaged in an arbitrary manner so long as the averaging does not
affect the background geometry or it need not be averaged at all.

Two useful methods are given for defining effective stress-energy
tensors.  One works well for 
stress-energy tensors that are averaged in some way and the other
works well when no averaging occurs.  For the former method it is
shown that there is much more freedom available than has usually
been thought in defining gauge invariant effective stress-energy
tensors for gravitational waves.  In the latter method there is
much more freedom than usually thought in defining the wave
equation for the gravitational waves.  In both cases this extra
freedom is due to the freedom available in separating the metric
into background and perturbed parts.

     In what follows the formalism used to describe gravitational
waves and other gravitational perturbations in spacetimes with no
classical matter is given first followed by the definition of the
generalized gauge transformations.  Then gauge invariance of the
perturbed Einstein tensor is established followed by a
description of the two methods of defining effective 
stress-energy tensors.  Approximate gauge invariance within the
context of an arbitrary perturbation expansion is next
established for stress-energy tensors that have been averaged in
some arbitrary way.  A method of solving the resulting wave and
backreaction equations is given.  The case of stress-energy
tensors which are not averaged is then considered.  Finally the
generalization to spacetimes containing classical matter is
discussed.

     To begin consider a separation of the metric into a
background part, $\gamma_{\mu \nu}$, and a perturbed part,
$h_{\mu \nu}$ such that  
\begin{equation}
    g_{\mu \nu} = \gamma_{\mu \nu} + h_{\mu \nu}
\end{equation}
The separation is arbitrary.  Following BH, the Einstein tensor
can similarly be divided into a part describing the curvature due
to the background geometry and that due to the perturbation by
writing
\begin{equation}
  G_{\mu \nu}(g) = G_{\mu \nu}(\gamma) + \bigtriangleup G_{\mu
\nu}(\gamma,h) = 0 \;\;\;.
\end{equation}
Here and hereafter indices of tensors are sometimes suppressed
for notational simplicity.  It is important to note that
$\bigtriangleup G$ is defined by this equation.  

     From Eq.(2) it is seen that the quantity $\bigtriangleup G$
is conserved with respect to the background geometry $\gamma$. 
It can also be seen that $\bigtriangleup G$ is invariant under
coordinate transformations which change the perturbed geometry
but leave the background geometry alone.  This property leads to
the invariance of $\bigtriangleup G$ under the generalized gauge
transformations which are defined below.  

Consider an arbitrary coordinate transformation.  It is always
possible to write such a transformation in the form
\begin{equation}
\bar{x}^\mu = x^\mu + \xi^\mu
\end{equation}
If the functional form of the background geometry is not allowed
to change under this coordinate transformation, that is if
$\bar{g} = \gamma + \bar{h}$, then $\bar{h}$ is given implicitly
by the equation
\begin{eqnarray}
\gamma_{\mu \nu}(x) + h_{\mu \nu}(x) &=& \gamma_{\mu
\nu}(\bar{x}) + \bar{h}_{\mu \nu}(\bar{x}) + (\gamma_{\mu
\alpha}(\bar{x}) + \bar{h}_{\mu \alpha}(\bar{x}))
{\xi^\alpha}_{,\nu} \nonumber \\
&   &  + (\gamma_{\alpha \nu}(\bar{x}) + \bar{h}_{\alpha
\nu}(\bar{x})) {\xi^\alpha}_{,\mu} + (\gamma_{\alpha
\beta}(\bar{x}) + \bar{h}_{\alpha \beta}(\bar{x}))
{\xi^\alpha}_{,\mu} {\xi^\beta}_{,\nu}
\end{eqnarray}
Here derivatives of $\xi$ are with respect to $x$ not $\bar{x}$.

A generalized gauge transformation is defined as one in which the
quantity $\bar{h}(x)$ is substituted for $h(x)$ into the
expression of interest.  If $h$, $\xi$ and their derivatives are
small enough, then to leading order this gauge transformation is
equivalent to the usual one used for gravitational waves which is
\begin{equation}
\bar{h}_{\mu\nu}(x) = h_{\mu\nu}(x) - \gamma_{\mu \alpha}(x)
{\xi^\alpha}_{,\nu} - \gamma_{\alpha \nu}(x) {\xi^\alpha}_{,\mu}
- \gamma_{\mu\nu,\alpha} \xi^\alpha \;\;\;.
\end{equation}

The quantity $\bigtriangleup G$ plays an important role in the
definitions of gravitational wave stress-energy tensors which
follow.  To prove its invariance under generalized gauge
transformations first note that, since the functional form of the
background metric is not to be changed by the coordinate
transformation (3),
\begin{equation}
\bar{G}(\bar{g}(\bar{x})) = G(\gamma(\bar{x})) + \bigtriangleup
\bar{G}(\gamma(\bar{x}),\bar{h}(\bar{x})) = 0 \;\;\;.
\end{equation}
Having obtained $\bar{G}(\gamma(\bar{x}),\bar{h}(\bar{x}))$ via a
coordinate transformation it is next useful to consider it simply
as a function of $\bar{x}$.  If this is done and it is evaluated
at $\bar{x} = x$ then combining Eq.(6) with Eq.(2) gives
\begin{equation}
\bigtriangleup \bar{G}(\gamma(x),\bar{h}(x)) = \bigtriangleup
G(\gamma(x),h(x)) \;\;\;.
\end{equation}

  If Eq.(4) is solved for $\bar{h}(x)$ as a function of $h(x)$
and this solution is substituted for $\bar{h}(x)$, then
$\bigtriangleup G$ is gauge invariant if it retains its original
functional form after these substitutions, that is if 
\begin{equation}
\bigtriangleup G(\gamma(x),\bar{h}(x)) = \bigtriangleup
G(\gamma(x),h(x)) \;\;\;.
\end{equation}  
To show that Eq.(8) is correct first note that the Einstein
tensor can be written in terms of a particular combination of the
metric tensor and its derivatives which is the same in any
coordinate system.  This implies that $\bigtriangleup \bar{G}$
can be computed by direct substitution of $\bar{h}$ into
$\bigtriangleup G$.  Thus
\begin{equation}
\bigtriangleup \bar{G}(\gamma(x),\bar{h}(x)) = \bigtriangleup
G(\gamma(x),\bar{h}(x)) \;\;\;. 
\end{equation}
Combining Eq.(7) and Eq.(9) gives Eq.(8).  This proves that
$\bigtriangleup G$ is invariant under gauge transformations of
the form $h(x) \rightarrow \bar{h}(x)$.

The next question to be addressed is the solution of Eq.(2).  It
is not possible to solve Eq.(2) directly without specifying in
some way the split between the background geometry and the
perturbed geometry.  One can of course either fix $\gamma$ and
solve for $h$ or vice versa.  However, to describe gravitational
waves the most useful methods are:
\begin{enumerate}
  \item  Define an effective stress-energy tensor for the
perturbed geometry which is conserved with respect to the
background geometry and invariant under generalized gauge
transformations, but which is otherwise arbitrary.  Then one can
write Eq.(2) as
     \begin{mathletters}
      \begin{eqnarray}
   \bigtriangleup G_{\mu\nu}(\gamma,h) &=& - 8 \pi
{T^{G}}_{\mu\nu}(\gamma,h) \\
  G_{\mu\nu}(\gamma) &=& 8 \pi {T^{G}}_{\mu\nu}(\gamma,h)
      \end{eqnarray}
     \end{mathletters}
  \item  Impose a gauge invariant equation which specifies the
perturbed geometry.  For gravitational waves it is the wave
equation.  It has the general form
  \begin{equation}
  H_{\mu\nu}(\gamma,h) = 0
  \end{equation}
In this case Eq.(2) is the backreaction equation and the
effective stress-energy tensor for the gravitational waves is
\begin{equation}
{T^{G}}_{\mu\nu}(\gamma,h) = - {1 \over {8 \pi}} \bigtriangleup
G_{\mu\nu}(\gamma,h)
\end{equation}
\end{enumerate}

     BH used Method (1) and made the following definition for
their effective stress-energy tensor:
\begin{equation}
{T^{G}}_{\mu\nu}(\gamma,h) = - {1 \over {8 \pi}} <\bigtriangleup
G_{\mu\nu}(\gamma,h)> \;\;\;.
\end{equation}
Here the angular brackets indicate a time average.  This was
useful for the problem they were considering which was the
gravitational geon.  Isaacson's definition is the same except
that the average is over a region of spacetime which is large in
scale compared to the wavelengths of the waves, but smaller in
size than the scale on which the background geometry varies.  The
definitions of BH and Isaacson can be extended to any averaging
procedure which does not affect $G(\gamma)$.  It is clear from
the above proof of the invariance of $\bigtriangleup G$ under
generalized gauge transformations that the BH stress-energy
tensor is gauge invariant.  

Once a gauge is chosen $<\bigtriangleup G>$ is still not
specified.  It can be set equal to any symmetric second rank
tensor which  does not depend on the variables which are averaged
over and is conserved with respect to the background geometry. 
This is because Eq.(10a) ensures that $\bigtriangleup G$ will
always have the correct average.  One interesting choice that can
be made is $<\bigtriangleup G> = 0$.  In this case the background
geometry is an exact solution to Einstein's equations and the
gravitational waves do not alter this geometry.  The choice
implicitly made by BH and Isaacson is discussed below. 

Method (2) can be used to define an effective 
stress-energy tensor without averaging.  One gauge invariant
choice for the wave equation is $H = \bigtriangleup G$.  This is
equivalent to using Method (1) and choosing $<\bigtriangleup G> =
0$.

     So far the discussion has been formal with exact results. 
If an expansion of the form
\begin{equation}
 \bigtriangleup G =  \bigtriangleup_1 G + \bigtriangleup_2 G +
... \;\;\;.
\end{equation}
exists then it can immediately be seen from Eqs.(2) and (14) that
to nth order it is the quantity $\bigtriangleup_1 G + ... +
\bigtriangleup_n G$ that is conserved with respect to the
background geometry.  Throughout when discussing the order of the
approximation, quantities whose value is of the same order as
$\bigtriangleup_n G$ are considered to be nth order quantities.  

Gauge invariance is more difficult.  Under an arbitrary gauge
transformation $h$ can change dramatically and in the new gauge
the appropriate perturbation expansion for $\bigtriangleup G$
might be very different than that in the original gauge.  This
would make it impossible to usefully compare terms in the two
expansions.  Thus when using perturbation expansions it is
usually necessary to restrict gauge transformations to those
which are small enough so that ${\bar{h}}$ is of the same order
of magnitude as $h$\cite{I}.  Then the perturbation expansions of
$\bigtriangleup G$ in both gauges are similar and can be
compared.  Therefore, throughout this paper, whenever a
perturbation expansion is used it is assumed that gauge
transformations are small enough so that
${\bar{h}}$ is of the same order of magnitude as $h$.   

     The proof of gauge invariance given above implies that
$\bigtriangleup_1 G + ... + \bigtriangleup_n G$ is gauge
invariant to nth order.  This means that, in general,
$\bigtriangleup_1 G$ by itself is only gauge invariant to first
order and $\bigtriangleup_2 G$ by itself is not gauge invariant
at all.  There are exceptions.  For the high frequency waves
considered by Isaacson, $\bigtriangleup_1 G$ and
$<\bigtriangleup_2 G>$ are both gauge invariant to second order. 

     If, for some arbitrary type of averaging, definition (13) is
used along with a perturbation expansion then Eqs.(10a,b) become
\begin{mathletters}
\begin{eqnarray}
\bigtriangleup_1 G + ... + \bigtriangleup_n G &=&
<\bigtriangleup_1 G + ... + \bigtriangleup_n G> \\  
G(\gamma) &=& -<\bigtriangleup_1 G + ... + \bigtriangleup_n G> 
\;\;\;.
\end{eqnarray}
\end{mathletters}
As discussed above it is still necessary, in a particular gauge,
to explicitly fix the right hand sides of these equations.  Once
this is done their general form assures gauge invariance to nth
order.  BH and Isaacson implicitly fix the values of their 
stress-energy tensors by imposing the condition
\begin{equation}
<\bigtriangleup_1 G> = 0
\end{equation}
This is a reasonable condition to impose because they are
considering high frequency gravitational waves and
$\bigtriangleup_1 G$ is linear in $h$ in that case.  

It is important to understand that the condition (16) (as opposed
to the stress-energy tensor itself) is only gauge invariant to
first order in general and even in Isaacson's case is only gauge
invariant to second order.  Thus when a transformation is made
from the original gauge where this condition is imposed to a new
gauge, care must be taken to determine the correct form of
$<\bigtriangleup_1 G>$ in the new gauge.  In general it no longer
vanishes.  It is this fact which makes it possible for the
combination $<\bigtriangleup_1 G + ... + \bigtriangleup_n G>$ to
be gauge invariant to nth order.

Isaacson developed a practical method of solving Eqs.(15a,b) for
the perturbation expansion that he used.  It is easily extended
to a general perturbation expansion.  One first expands $h$ in
the following way:
\begin{equation}
h = {h^{(1)}} + {h^{(2)}} + ...\;\;\;.
\end{equation}
Here the terms on the right hand side are defined such that
$\bigtriangleup_1 G(\gamma,h^{(2)})$ is of the same order as
$\bigtriangleup_2 G(\gamma,h^{(1)})$, and so forth.  Then to
second order Eqs.(15a,b) can be written
\begin{mathletters}
\begin{eqnarray}
\bigtriangleup_1 G(\gamma,h^{(1)}) &=& <\bigtriangleup_1
G(\gamma,h^{(1)})> \;\;\;, \\
\bigtriangleup_1 G(\gamma,h^{(2)}) + \bigtriangleup_2
G(\gamma,h^{(1)}) &=& <\bigtriangleup_1 G(\gamma,h^{(2)}) +
\bigtriangleup_2 G(\gamma,h^{(1)})> \\
G(\gamma) &=& - <\bigtriangleup_1 G(\gamma,h^{(1)}) +
\bigtriangleup_1 G(\gamma,h^{(2)}) + \bigtriangleup_2
G(\gamma,h^{(1)})> 
\end{eqnarray}
\end{mathletters}
The extension to higher orders is straight-forward.  This method,
along with a condition which fixes the stress-energy tensor in a
particular gauge, results in a consistent set of equations that
can be solved.  

As an example, consider the case of gravitational waves which
have small amplitudes, frequencies and momenta.  For these waves
the Einstein tensor can be expanded in powers of $h$ and its
derivatives with the result that $\bigtriangleup_1 G(\gamma,h)$
is first order in $h$, $\bigtriangleup_2 G(\gamma,h)$ is second
order, and so forth\cite{BH,I}.  The perturbed metric is expanded
as in Eq.(17) and the equations to second order are given in
(18a-c).  Since $\bigtriangleup_1 G$ is linear in $h$, Eq.(18b)
implies that ${{h}^{(2)}}$ is of order $({{h}^{(1)}})^2$.  The
condition (16) can be imposed by requiring that
$<\bigtriangleup_1 G(\gamma,h^{(n)})> = 0$ for all $n$.

     Method (2) results in a gauge invariant effective 
stress-energy tensor for gravitational waves when no averaging
occurs.  If a perturbation expansion of the form (14) is used it
is natural to define the approximate wave equation to be
$\bigtriangleup_1 G = 0$.  Unfortunately this does not lead to a
gauge invariant stress-energy tensor.  However if the expansion
(17) is used then the wave equation can be defined to be
\begin{mathletters}
\begin{equation}
\bigtriangleup_1 G(\gamma,h^{(1)}) = 0 \;\;\;.
\end{equation}
The resulting backreaction equation to second order is
\begin{equation}
G(\gamma) = - \bigtriangleup_1 G(\gamma,h^{(2)}) -
\bigtriangleup_2 G(\gamma,h^{(1)}) \;\;\;.
\end{equation}
\end{mathletters}
A proof similar to Isaacson's proof of the approximate gauge
invariance of the wave equation shows that, in this case, the
wave and backreaction equations are gauge invariant to second
order.  The stress-energy tensor is also conserved with respect
to the background geometry to second order.  Thus the wave and
backreaction equations are consistent to this order.

     It is useful to generalize the above results to the case in
which matter is present\cite{footnote}.  If the matter fields can
be described by a covariant action then both the wave equation
and the 
stress-energy tensor for the matter can, in any coordinate
system, be written as some particular combination of the matter
fields, the metric tensor and their derivatives.  This makes it
possible to split the wave equation and the stress-energy tensor
into background and perturbed parts just as was done for the
Einstein tensor in Eq.(2).  If the matter fields are denoted by
$\phi$, the wave equation by $W$, and the stress-energy tensor
for the matter by $T$, then the wave equation and Einstein's
equations can be written as
\begin{mathletters}
\begin{eqnarray}
W(\gamma,\phi) &=& - \bigtriangleup W(\gamma,h,\phi) \\
G(\gamma) &=&  8 \pi T(\gamma,\phi) + 8 \pi
\bigtriangleup T(\gamma,h,\phi) - \bigtriangleup G(\gamma,h)
\;\;\;.
\end{eqnarray}
\end{mathletters}

   From Eq.(20b) it is clear that the combination $\bigtriangleup
G(\gamma,h) - 8 \pi \bigtriangleup T(\gamma,h,\phi)$ is conserved
with respect to the background geometry.  Two proofs of exactly
the same type as that establishing the gauge invariance of
$\bigtriangleup G$ when no matter is present, show that the
quantities
$\bigtriangleup W(\gamma,h,\phi)$ and 
$\bigtriangleup G(\gamma,h) - 8 \pi \bigtriangleup
T(\gamma,h,\phi)$ are invariant under generalized gauge
transformations. 
 
It is not difficult to show that all of the results found for the
vacuum case go over in a straight-forward manner to the case when
matter is present if the substitution $\bigtriangleup G
\rightarrow \bigtriangleup G - 8 \pi \bigtriangleup T$ is made. 
Thus regardless of whether or not averaging occurs it is possible
to derive a self-consistent set of equations which describe the
behavior of the gravitational waves and the matter fields as well
as their effects upon the background geometry. 

     In conclusion, use of the generalized gauge transformation
implicitly given by Eq.(4) makes it possible to virtually always
define an effective stress-energy tensor for gravitational waves
and other perturbations which is both conserved and gauge
invariant.  The stress-energy tensor may be averaged in some
arbitrary manner that does not affect the background geometry or
it need not be averaged at all.  

     I would like to thank D. Brill and J. Hartle for very
helpful discussions.  This work was supported in part by Grant
Number PHY95-12686 from the National Science Foundation.


\begin{references}
\bibitem{footnote} This includes the case of a nonzero
cosmological constant.  The effective stress-energy tensor for
the cosmological constant is $T_{\mu\nu} = - g_{\mu\nu} \Lambda/8
\pi$.
\end{references}
\end{document}